\documentstyle[11pt,aaspp,psfig]{article}
\def\gs{\mathrel{\raise0.35ex\hbox{$\scriptstyle >$}\kern-0.6em % Greater/squiggles
\lower0.40ex\hbox{{$\scriptstyle \sim$}}}}
\def\ls{\mathrel{\raise0.35ex\hbox{$\scriptstyle <$}\kern-0.6em % Less than/squiggles
\lower0.40ex\hbox{{$\scriptstyle \sim$}}}}

\def\arcsper{\ifmmode \rlap.{''}\else $\rlap{.}''$\fi}
\def\arcmper{\ifmmode \rlap.{'}\else $\rlap{.}'$\fi}

\def\UV{\hbox{$(U\! -\! V)$}}
\def\dUV{\hbox{$\delta(U\! -\! V)$}}
\def\VVII{\hbox{$(V_{555}\! -\! I_{814})$}}

\def\VI{\hbox{$(V\! -\! I)$}}

\def\II{\hbox{$I_{814}$}}
\def\VV{\hbox{$V_{555}$}}
\received{---}
\revised{---}
\accepted{---}

\begin{document}
\small
%\doublespace

\title{THE HOMOGENEITY OF SPHEROIDAL POPULATIONS IN DISTANT CLUSTERS\footnote{
Based on observations 
obtained with the NASA/ESA Hubble Space Telescope
which is operated by STSCI for the Association of Universities
for Research in Astronomy, Inc., under NASA contract NAS5-26555.}}
\author{
Richard S.\ Ellis\altaffilmark{1},
Ian Smail \altaffilmark{2}\footnote{Visiting Research Associate at the Carnegie
Observatories.},
Alan Dressler\altaffilmark{3},
Warrick J.\ Couch\altaffilmark{4},\\
Augustus Oemler Jr.\altaffilmark{5}\footnote{Present address: 
 The Observatories of the Carnegie Institution of Washington, 813 Santa Barbara St., Pasadena, CA 91101-1292},
Harvey Butcher\altaffilmark{6} \& 
Ray M.\ Sharples\altaffilmark{2},
}

\affil{1) Institute of Astronomy, Madingley Rd, Cambridge CB3 OHA, UK}
\affil{2) Department of Physics, University of Durham, South Rd, Durham DH1 3LE, UK}
\affil{3) The Observatories of the Carnegie Institution of Washington, 813 Santa Barbara St., Pasadena, CA 91101-1292}
\affil{4) School of Physics, University of New South Wales, Sydney 2052, Australia}
\affil{5) Astronomy Department, Yale University, PO Box
208101, New Haven CT 06520-8101}
\affil{6) NFRA, PO Box 2, NL-7990, AA Dwingeloo, The Netherlands}

%%\vfil\eject

\begin{abstract}
The small scatter observed for the \UV\ colors of spheroidal galaxies
in nearby clusters of galaxies provides a powerful constraint on the
history of star formation in dense environments. However, with local
data alone, it is not possible to separate models where galaxies
assembled synchronously over redshifts $0 < z < 1$ from ones where
galaxies formed stochastically at much earlier times. Here we attempt
to resolve this ambiguity via high precision rest-frame UV--optical
photometry of a large sample of morphologically-selected spheroidal
galaxies in three $z\sim0.54$ clusters which have been observed with
HST. We demonstrate the robustness of using HST to conduct the
morphological separation of spheroidal and disk galaxies at this
redshift and use our new data to repeat the analysis conducted locally
at a significant look-back time.  We find a small scatter ($< 0.1$ mag
rms) for galaxies classed as Es and E/S0s, both internally within each
of the three clusters and externally from cluster to cluster. We do not
find any trend for the scatter to increase with decreasing luminosity
down to $L\sim L_V^*+3$, other than can be accounted for by
observational error. Neither is there evidence for a distinction
between the scatter observed for galaxies classified as ellipticals and
S0. Our result provides a new constraint on the star formation history
of cluster spheroidals prior to $z\simeq0.5$ confirming and
considerably strengthening the earlier conclusions. Most of the star
formation in the elliptical galaxies in dense clusters was
completed before $z\simeq3$ in conventional cosmologies. Although we
cannot rule out the continued production of {\it some} ellipticals, our
results do indicate an era of initial star formation consistent with
the population of star-forming galaxies recently detected beyond
$z\simeq3$.
\end{abstract}


\sluginfo

%\vfil\eject

\section{Introduction}

Elliptical galaxies are conventionally regarded as old galactic systems
whose star formation history can be approximated as a single burst that
occurred 12--16 Gyr ago (Baade 1958; Tinsley \& Gunn 1976; Bruzual
1983). However, in recent years, this simple picture has been
subjected to a number of challenges. Numerous cases have been found of
ellipticals with intermediate-age stellar populations (O'Connell 1980)
and dynamical arguments suggest that many peculiarities seen in
ellipticals (shells and dust-lanes) are best explained via recent
formation from the merger of gas-rich systems (Toomre 1978; Quinn
1984). Detailed spectroscopy of such systems provides a rich set of
data whose interpretation in terms of star formation history requires
an adequate separation of the competing effects of metallicity and age
(Worthey 1995).

The conflict might be resolved if the majority of ellipticals were old single
burst systems, whereas the remainder formed via merging of gas-rich disk
galaxies. In this case one might expect an environmental and/or mass
dependence in the rate of occurrence of intermediate age populations.
Reasonably good evidence is emerging that recent star formation is more
prevalent in low density environments than in clusters. Rose et al.\ 
(1994) find the mean stellar dwarf/giant ratio is higher in environments
with low virial temperatures. This would be consistent with other
environmental trends which indicate accelerated star formation histories
in clusters (Oemler 1991). 

Sandage \& Visvanathan (1978) first proposed that the UV-optical
color-magnitude (C-M) relation for cluster galaxies could provide a
significant constraint on their past history of star formation. The
improved precision possible with modern CCD detectors was exploited by
Bower, Lucey \& Ellis (1992, hereafter BLE) who obtained high precision
$U$ and $V$ photometry of spheroidal galaxies in two local clusters,
Virgo and Coma. BLE observed a very small scatter, \dUV$ \ls 0.035$,
around the mean color-luminosity relation for luminous E/S0's, which
was barely larger than that attributable to observational errors.
 
The sensitivity of the $U$-band light to small numbers of hot, young
stars enabled BLE to constrain the past contribution from upper main
sequence stars statistically across both cluster samples. With a single
present-day value for the scatter they chose to express their result in
a number of ways.  Firstly, in terms of the passive evolution of a
single-burst population, the tightness of the C-M relation was used to
derive a minimum age.  In the absence of any recent star-formation, the
homogeneity of the Virgo and Coma populations would suggest stochastic
formation of galaxies could only occur within the first few Gyr after
the big bang (i.e.\ $z>$2 in cosmologies with $H_o$=50 and
$\Omega$=0.1). Unfortunately, such absolute age estimates are subject
to many of the uncertainties which afflict those for globular
clusters.

More convincingly, abandoning the constraint on the age of the first
burst, BLE also concluded that no more than 10\% of the current stellar
population in present-day E/S0s could have been formed in any subsequent
activity in the past 5 Gyr as might be the case if merging of gas-rich
systems had been involved. They expressed this result more generally in
terms of a combined synchronicity+age constraint. Recent activity from
major merging would only be consistent with the tight \UV\ scatter if the star
formation history had been similar from galaxy to galaxy both within
each cluster and between the Coma and Virgo clusters.

BLE's result presents an important challenge for bottom-up hierarchical
theories of structure since these predict relatively recent formation
eras for massive galaxies.  Baugh et al.\ (1996) and Kauffmann (1996)
have addressed the question quantitatively using a simple prescription
for merger-induced star formation. They find that the homogeneity of
BLE's C-M data can be satisfied if the merging of disk galaxies that
produce the spheroidals was largely complete by a redshift $z\simeq$0.5. 

Although good progress has been made in tracking the UV-optical
C-M relation to higher redshift (Ellis et al.\ 1985; Arag\'on-Salamanca et
al.\ 1991, 1993), without morphological information a major uncertainty
remains. The scatter of the photometric C-M relation may be
underestimated if some spheroidal galaxies lie blueward of the C-M
sequence. This could well be the case if the timescale for dynamical
evolution is shorter than that for main sequence evolution as indicated
in numerical simulations (Mihos 1995; Barger et al.\ 1996a). This point
was originally examined by MacLaren et al.\ (1988) who showed that a
detectable UV excess could remain for $\simeq$2 Gyr {\it after} a burst
induced, for example, by a merger.

In this paper, we extend the analysis of BLE to a sample of three
$z\simeq$0.54 clusters, taking advantage of images from the {\it Hubble
Space Telescope} (HST) Wide Field Planetary Camera 2 (WFPC-2) to
morphologically classify a sample of faint galaxies. In \S2 we
summarise the new observations and describe our photometric techniques
and morphological classifications.  A key question we examine is the
reliability of using HST to isolate distant spheroidals. In \S3 we
discuss the color-magnitude relationships, both absolutely in terms of
predictions for no evolution and in terms of the photometric scatter.
\S4 discusses the main result in the context of the star formation
history of cluster galaxies and explores the likely consequences in
terms of locating the epoch of their formation.  \S5 summarises our
principal conclusions.
 
\section{Data}

\subsection{Observations}

The observations presented here consist of deep F555W and F814W images
of the core regions of three $z\simeq$0.54 clusters obtained using the
WFPC-2 (Figure~1).  The data form part of a larger study of the
morphologies of galaxies in distant clusters (the ``MORPHS'' project --
Smail et al.\ 1996a,b (S96); Dressler et al.\ 1996, 1997; Barger et
al.\ 1996b). The clusters discussed here cover a range of optical
richnesses and X-ray luminosities within a narrow redshift interval
specifically selected so that observed F555W/F814W is close to rest-frame
$U$/$V$.  At $z=0.54$ the F555W passband ($525\pm 61$ nm) samples $341 \pm
40$ in the restframe with F814W ($828\pm 88$ nm) lying at $538 \pm 57$
nm, compared to $U$ and $V$ at $365\pm 70$ nm and $550 \pm 90$ nm
respectively.  Using these passbands we can thus analyse our cluster
data in a similar way to BLE, the principal difference being the
significant look-back time. In order of decreasing mass or X-ray
luminosity, the clusters are Cl0016+16 at $z=0.546$ (Koo 1981; Ellis et
al.\ 1985; Dressler \& Gunn 1992; Arag\'on-Salamanca et al.\ 1993),
Cl0054$-$27 at $z=0.563$ (Couch et al.\ 1985, 1991) and Cl0412$-$65
($\equiv F1557.19TC$) at $z=0.510$ (Couch et al.\ 1991; Bower et
al.\ 1994, 1996). Further properties of these systems and a log of the
HST observations are given in Table~1.

The individual exposures in each passband are grouped in 2 sets, each
offset by 2 arcsec to allow hot pixel rejection.  After standard
pipeline reduction, the images were aligned using integer pixel shifts
and then combined into final F555W/F814W frames using the IRAF/STSDAS
task CRREJ.  We chose to work in the WFPC-2 filter system (which we
call \VV\ and \II), calibrated from our own ground-based images of the
clusters using the color transformations given in Holtzman et
al.\ (1995), as discussed below.  The final images (Figure~1) cover the
central 0.7 h$^{-1}$ Mpc,\footnote{Unless otherwise stated, we use
$q_o=0.5$ and $h = H_o / 100$ km sec$^{-1}$ Mpc$^{-1}$.  Thus 1 arcsec
$\equiv$3.6 h$^{-1}$ kpc (Cl0412$-$65) or 3.7 h$^{-1}$ kpc (Cl0016+16,
Cl0054$-$27)} i.e.\ $\sim 0.35$ h$^{-1}$ kpc/pixel.  The processed data
reaches a 5$\sigma$ limiting depth of $I_{814}=26.7$ and provides
\VVII\ colors with better than 2\% precision at $I_{814}=21.0$ and 5\%
at the limit of $I_{814}=23.0$ used in this analysis (see below).

As we will compare this analysis with BLE's previous study of
early-type galaxies in the Coma and Virgo clusters, it is important to
consider the physical regions of the clusters sampled as well as the
spatial resolution and photometric precision of the relevant
ground-based and HST datasets. In the case of BLE, galaxies were
selected according to the availability of stellar velocity dispersions
from Dressler (1984); generally they lie within 500 $h^{-1}$ kpc of the
cluster centre. BLE's analysis was based on aperture photometry within
a diameter $\simeq$5$h^{-1}$ kpc (60 arcsec at Virgo, 11 arcsec at
Coma). Over a total luminosity range of $\simeq$4 mags to a limiting
absolute magnitude of $M_V \simeq -17 + 5 \log h$, the rms photometric
error in \UV\ was 0.03 mag. As we will see below, we can reproduce these
quantities fairly closely with the HST samples.

\subsection{Reduction and Photometric Precision}

We selected galaxies in the HST frames from the stacked F814W
(restframe $\sim V$) images using the SExtractor package of Bertin \&
Arnouts (1996).  After convolving the data with a top-hat kernel of
diameter 0.3 arcsec, all objects with areas greater than 12 pixels
above the $\mu_{814} = 25.0$ mag arcsec$^{-2}$ isophote were evaluated.
This ensures a source list which extends much fainter than that
eventually used in this analysis.  Color were measured using an
aperture with a diameter of 5 h$^{-1}$ kpc ($\sim 1.4$ arcsec) closely
matching that used by BLE and total magnitudes were obtained from the
SExtractor {\bf BEST\_MAG} estimates. We discuss the precision of both
the aperture and integrated photometry next.

\subsection{Calibration of \VVII\ between clusters}

An important aim of our investigation is to not only examine the
scatter {\it internal} to each cluster but also {\it externally} across
the three clusters.  This requires precise relative colors between the
three clusters and we have thus chosen to independently determine the
photometric zero-points of our WFPC-2 images using ground-based images
of the HST fields, rather than having to assume that our various WFPC-2
F555W/F814W observations have fixed zero-points.  The ground-based data
were all obtained on a single photometric night and thus provide a
robust calibration of the colors of galaxies across the three
clusters.  The images were kindly provided by Barry Madore and were
taken using a thinned 2k$^2$ Tek CCD on the 2.5m Du Pont telescope at
Las Campanas Observatory (LCO), Chile.  The clusters were observed in
Johnson $V$ and Cousins $I$ across a range in airmasses, 1.2--1.5, on
the night of October 14th, 1996.  A large number of Landolt (1992)
standard fields were also observed during this time, bracketing the
airmass range of our observations.  The total integrations on our
clusters are 500s in $V$ and 800s in $I$, with seeing of FWHM=1.0--1.2
arcsec.  These data were reduced and calibrated in a standard manner
using dome and sky flats.    Analysis of the standard fields shows that
the night was photometric and allows us to derive extinction and color
calibration of our science observations to an accuracy of $\pm 0.029$
in $V$ and $\pm 0.015$ in $I$, where we include the uncertainty for
extrapolating the color correction from the range of our standard stars
$(V-I)\sim 1$ to the expected colors of E/S0's in the distant clusters
$(V-I)\sim 2.5$.  

We next measure photometry within 3 arcsec diameter apertures from
these $V$ and $I$ images, having matched the seeing in the frames from
measurements of the stellar profiles.  To reduce our sensitivity to
residual color corrections we have chosen to restrict our comparison to
just those red E/S0 galaxies from our HST images which are used in the
analysis below.  We also apply a further magnitude limit to this sample
owing to the shallower depth of our ground-based images, hence we
restrict ourselves to galaxies with $I\leq 20.5$.  Having measured the
colors of these galaxies on the ground-based images we convert our
\VI\ to \VVII\ using the color equations given for the WFPC-2 flight
system in Table~7 of Holtzman et al.\ (1995).  We can now compare this
photometry with that obtained from our WFPC-2 images, where we have
assumed a single zero-point for the F555W and F814W images of the three
clusters.  We determine offsets in the zero-points of the WFPC-2
photometry relative to the LCO data of $-0.017\pm 0.035$, $-0.062\pm
0.043$ and $0.072\pm 0.036$ for Cl0016+16, Cl0054$-$27 and Cl0412$-$65,
where the errors in the means are estimated by bootstrap resampling.
While using a single WFPC-2 zero point would be accurate in the mean:
$-0.002\pm  0.068$, the scatter between the clusters (especially for
Cl0412$-$65) justifies our recalibration.   Having applied these
corrections to our photometry we now have a robust calibration of 
\VVII\ across our three clusters with a precision of 4\%.

\subsection{Calibration of \VVII\ between WFC CCDs}

We have also tested the relative photometry across the WFPC-2 field to
estimate contributions to the scatter of the entire sample from
possible zero-point variations between the CCD chips.  Unfortunately,
the calibrations shown above lack sufficient numbers of objects to
allow us to undertake this test for all three clusters.  However, we
can apply this with adequate precision by comparing our HST photometry
with deep high resolution ground-based images of a single cluster,
Cl0016+16 (Smail 1993; Smail et al.\ 1994), to search for systematic
offsets as a function of position.  Here we are searching for a
variation in the difference between the colors of galaxies measured on
the ground-based image and the three WFC CCDs.   Owing to the slightly
different color responses of the ground-based Johnson $V$ and Cousins
$I$ compared to F555W and F814W we compare photometry only for those
objects which lie in a relatively narrow band in the HST
color-magnitude diagrams.  To \II $<$23.0 we have $\sim 230$ galaxies
across the three chips. We set a firm upper limit of $\Delta = 0.02$
mag on the relative offsets in \VVII\ between the chips.

\subsection{Photometric precision}

The photometric errors on our HST aperture magnitudes have been estimated
by analysing both the variance in the background sky and the poisson
noise in the object counts. We have confirmed the estimates so obtained
by splitting the imaging data for one of the clusters (Cl0016+16) into
two independent sets and comparing the photometry of 160 of the \II
$<$23.0 objects measured off each half. For \II =22.0 we find an
average photometric error in \VVII\ of $\pm$0.016 increasing to
$\pm$0.063 at \II = 23.0; the median values are $\pm$0.015 and
$\pm$0.053 respectively.  The distribution of the differences for the
two independent datasets in Cl0016+16, when scaled by the theoretical
uncertainties shows a distribution consistent with these errors. The
typical difference in integrated \II\  magnitudes from the SExtractor
photometry of the Las Campanas and WFPC-2 images of the spheroidal
galaxies in the three clusters is $(I_{814}^{\rm HST}-I_{814}^{\rm
LCO}) \simeq 0.1$ mag at \II$\simeq$21.0. Examining the variance in the
profile fits, we also estimate an uncertainty of $\pm 0.1$ mag in the
total magnitudes adopted.

We will generally consider three magnitude-limited samples for each of
our clusters. The brightest sample is limited at \II $<21.0$
corresponding to an absolute magnitude of $M_V \simeq -19.8 + 5 \log h$
($L_V \sim L^\ast +1$).  At this limit the morphological
distinction between Es and S0s is relatively clear, although we discuss below
the level of mixing between these two samples due to misclassification.
To \II $=22.0$, a large fraction of the galaxies are
spectroscopic-confirmed members and the photometric precision is
similar to that obtained by BLE.  The faintest sample is limited at \II
$<23.0$ -- or $M_V < -17.8 + 5 \log h$ -- and enables us to investigate
the possibility of a variation in the \VVII\ scatter along the cluster
C-M relation for the spheroidal (E and S0) population as a whole.

Finally, we compare the precision of the photometry available from our
HST images and the 4.2m William Herschel Telescope (Smail 1993; Smail
et al.\ 1994) for Cl0016+16.  At the faintest limit of our
classifications, \II =23.0, the HST rms error on \VVII\ is 70\% of that
obtained from the larger aperture WHT, despite good conditions and
longer integrations.  This can be understood in terms of the reduced
sky background at longer wavelengths. Specifically, the \II\ sky is
$\simeq$8 times fainter in space and this gain is equivalent to an
effective aperture increase of $\simeq\times$2.6 for background-limited
work. Thus HST has significant photometric advantages at faint limits
quite apart from its superior image quality.

\subsection{Photometric Corrections}
 
Before analysing the C-M relations and the associated UV--optical
scatter, it is important to introduce two further small corrections
that must be made to the galaxy photometry if we wish to combine the
data across all three clusters.

Firstly, we have taken the interstellar reddening for our three fields
from the NED database.\footnote{NED (the NASA/IPAC Extragalactic
Database) is operated by the Jet Propulsion Laboratory, California
Institute of Technology, under contract with the National Aeronautics
and Space Administration.} We find $A_B$ = 0.08, 0.06 and 0.10 for
Cl0016+16, Cl0054$-$27 and Cl0412$-$65 respectively. As these effects
are differentially very small, we will apply corrections only to the
\VVII\ colours using the conversion $\delta\,\VVII$= 0.36 $A_B$ and
reducing all colors to those appropriate for Cl0016+16. The corrections
involved are small ($<$0.007).
 
Secondly, a small correction must be made to allow for the relative
$k$-correction difference across the three clusters, which have
slightly different redshifts. This can be done most simply by adopting
a spectral energy distribution (SED) and taking the derivative
$\delta\,k(V_{555}-I_{814}) / \delta\,z$ at the mean cluster redshift
$\overline{z}$=0.54. Adopting the SED of a present-day giant elliptical
with $M_V\simeq -19.8 + 5\log h$ ($I_{814}\sim 21$ at $z=0.54$)
(Arag\'on-Salamanca et al.\ 1993) gives $\delta\,k(V_{555}-I_{814})$ /
$\delta\,z$=1.8 and thus the differential color shift is found to be
$<$0.05 mag. 

As the slope of the C-M relation is relatively constant with redshift
(see below), provided data is compared from one cluster to another
at the same luminosity, the second-order effect to the redshift correction 
arising from SED variations with luminosity is very small. This was 
examined by interpolating to the SED of a present-day Sab galaxy (c.f.\ techniques discussed in Couch et al.\ 1984) where we found a negligible 
difference in the differential color term ($<$0.01).  

Combining both reddening and redshift effects, the color corrections
we have applied are $-$0.025 and 0.057 for Cl0054$-$27 and 
Cl0412$-$65, respectively. 

\subsection{Morphological Classification}

A central issue to an application of the BLE technique at
$z\simeq$0.54 is the separation between disk and spheroidal galaxies
from the HST images. In this section we describe our classification
methods and various tests conducted to test the reliability of our
results.

The morphological classifications for the various samples in each of
the three clusters were first determined independently by three of us
(AD/WJC/RSE) and subsequently combined into a single class according to
a scheme we have defined as part of the larger project to study the
evolution of galaxies in our cluster sample (S96).  The images were
classified blind according to a list with no regard to either the
spectroscopic or color data. More extensive details of the
classification scheme and its precision are given in our catalog paper
(S96).  Here we are primarily concerned with the accurate selection of
the spheroidal population and the precision with which the E and S0
subclasses can be identified to various depths.  In our classifications
we use the term `E/S0' to describe an ambiguous case, E or S0, rather
than to classify actual transition cases (in this we differ from the
approach used in the Revised Hubble scheme).  
 
The suitability of $\simeq$3-6 orbit WFPC-2 data (shallower than that
used here) for the morphological classification of distant galaxies has
been addressed by Glazebrook et al.\ (1995) and Abraham et al.\ (1996a)
by simulating the appearance of $z\simeq$0.7 galaxies in the F814W band
using nearby CCD data. Those studies carefully take into account the
various detector and imaging point spread functions and, in the case of
Abraham et al.\ (1996b), accurately allow for $k$-correction losses on
a pixel-by-pixel basis using spectral energy distributions assigned
from multicolor data.  Those workers found that misclassifications
generally only become significant beyond $z\simeq$1 and then mostly for
intermediate and late-type spirals. The combination of the differential
$k$-correction between disk and bulge light and surface brightness
losses imply a shift to apparently later types.

In this study a key question is the reliability of isolating the
boundary between early-type spirals and S0s as well as the distinction
between Es and S0s, which is troublesome even at low redshift.  One
worry is that, even over a smaller range in redshift for which the
differential $k$-correction effect is small, surface brightness losses
may make certain early spirals appear more bulge-dominated. That such
effects are unlikely to be significant is illustrated by Figure~2
(kindly prepared by Roberto Abraham) which presents a mosaic of
representative present-day elliptical, S0 and Sa galaxies taken from the
catalog of Frei et al.\ (1996) simulated as they would appear in the F814W
band at $z$=0.5 taking into account all the various HST-specific factors
discussed above.  Clearly, the majority of the diagnostic features used to
differentiate these systems locally remain clearly visible.

Turning to the cluster images, Figure~3 illustrates both the ease with 
which the classifications can be made to various limits as well as 
highlighting some difficult cases we encountered. Where possible we have 
chosen spectroscopically confirmed members. A comparison of the 
morphological classifications of the three classifiers is our primary guide 
since we wish the classifications to be strictly independent of color 
and spectroscopic properties.

To \II=21.0 across all Hubble types, the morphological types on a
scheme of 10 classes (S96) are in unanimous agreement for the three
classifiers for 70\% of the galaxies and within $\pm$1 type for all
objects (S96).  Most significantly for this study, 86\% of the
ellipticals and 60\% of the S0s to \II=21.0 were unanimously chosen as
such by all three classifiers. Examples of these objects are contained
in the top two rows of Figure~3. Note that, in calculating this success
rate, instances where one classifier indicated `E/S0' have been
rejected. Together with Abraham's simulations, this suggests the
morphological types to \II=21.0 are as good as those in the local
clusters, in particular for discriminating between spheroidals and disk
galaxies.

~From \II=21.0 to 23.0 the overall agreement across the Hubble sequence
worsens as might be expected. Examples of these galaxies are shown in
the third row of Figure~3. Unanimous agreement amongst the three
classifiers across all types falls from 70\% to 40\% with about 90\% of
the objects being classed to within $\pm$1 type (previously 100\%).
Fortunately, this does not impact significantly on the precision with
which ellipticals can be identified. 70\% of the classed Es were chosen
definitely as such by all three classifiers. However only 25\% of the
S0s were similarly identified.  

The issue of misclassification of round E and S0 galaxies is dicussed
in detail in S96, but we review it here briefly owing to its importance for this
analysis.  Discriminating between E and face-on S0 galaxies has always
been a concern for morphological classification.  A crude estimate of
the proportion of misclassified face-on S0s can be made from the
observed ellipticity distribution of S0s, under the assumption that
they can be simply modelled as randomly-projected thin disks.  Such a
model predicts a flat distribution of numbers of S0's with ellipticity,
when compared to this the observed fraction of S0s provides an estimate
on the likely misclassification rate.  Looking at the fraction of S0s
with $\epsilon \leq 0.3$ we find $23\pm 10$\% to \II =23 in the three
clusters.  Compared to the expected fraction of 33\% (S96) this
indicates that we maybe misclassifying up to 10\% of the S0
population.  When combining all the data from our three clusters (22 S0
galaxies) this amounts to 2--3 galaxies in total (compared to 60-or-so
ellipticals) and so it is difficult to see how it might strongly affect
our results.  We will incorporate this estimate into our analysis below, as
well as other possible sources of morphological contamination.  

Finally, in addition to the visual classification, we investigated a more
quantitative approach based on surface photometry profiles from the HST
images. We considered this might be helpful in addressing the
possibility that some bulge-dominated systems with faint disks may be
misclassified as ellipticals. We examined surface brightness profiles
for 12 cases with \II=21.0--22.0 where there was some disagreement
between the morphological classifiers. For five galaxies convincingly
identified to be ellipticals on the basis of profiles, four had been
identified as such morphologically by two of the three classifiers but
the third disputing classifier led to our being cautious and assigning
E/S0 in two cases. Only one profile-classed elliptical was thought
morphologically to be a S0. Likewise, two disk galaxies identified from
profiles had been assigned E/S0 on the basis of the morphologies
suggesting a similar effect. For the remaining five cases where there
was disagreement between the morphologists, the profile data was
inconclusive. The results suggest that whilst there is no major
discrepancy between the two approaches to classification, the surface
brightness profiles are not particularly helpful in deciding in cases
of difficulty. 

In summary therefore, the morphological selection of spheroidal
galaxies appears reliable to the limit of our sample. However, the
distinction between E and S0 becomes somewhat uncertain fainter than
\II=21.0--22.0 and may result in upto 10\% of our S0s being
misclassified as E.
 
\section{Analysis}

\subsection{Estimating the Field Contamination}

It is important at this stage to consider the possibility of
contamination of the cluster data from field galaxies, especially
any that might be morphologically early-type galaxies. First we discuss
the limited spectroscopic data that is available for our clusters.

Cl0016+16 has been studied by Dressler and Gunn (1992) and also, more
recently, by the CNOC group (Carlberg et al.\ 1996) who have kindly
made their results available to the authors (Abraham, private
communication).  Excluding stars, the original Dressler and Gunn
spectroscopy provides redshifts for 31 galaxies in the WFPC-2 field of
which 7 are non-members. The CNOC group observed 14 galaxies in the
same area of which 8 have redshifts in close agreement with Dressler
and Gunn; 6 are new redshifts and all are cluster members.   Examining
the 7 non-members, the redshifts cover a wide range and none are
morphologically spheroidal galaxies. The foreground group at $z$=0.30
suggested by Ellis et al.\ (1985) on the basis of their SED
fitting does not feature as prominently as those authors had suggested.
Only three galaxies have redshifts associated with this structure. 

In Cl0412$-$65, spectra are available for our new study (Dressler et
al.\ 1997) for 4 new members and a further 4 non-members.  In
addition spectra exist for 13 more galaxies from the recent study of
Bower et al.\ (1996) of which 5 are non-members. Amongst the 9
spectroscopically-confirmed field galaxies, none was classed as
spheroidals and thus no correction for contamination can be made on
this basis.  For Cl0054$-$27 spectroscopy is published for only the two
brightest members (Couch et al.\ 1985) and we supplement this with new
data on a further 10 members and 4 non-members from Dressler et
al.\ (1997).  Again there were no field spheroidals identified.

Overall, the spectroscopic coverage is sufficiently poor, that
field contamination must therefore be evaluated statistically.
Glazebrook et al.\ (1995) and Abraham et al.\ (1996b) present the
morphologically-dependent counts from the HST Medium Deep Survey and as
one of us (RSE) was involved in the classification of that sample, it
is well suited for estimating the number of field E/S0s in each cluster
field as well as their color distribution. As the MDS data has only
been classed to \II=22.0, we have also taken the deeper classifications
of Driver et al.\ (1995) as well as counts from the Hubble Deep Field
(Abraham et al.\ 1996a). These data indicate that we can expect at most
$\simeq$5-7 field E/S0s per cluster image to \II=23.0 whose colors
mostly lie in the range 1$<\VVII<$2. The uncertainty in subtracting the
field color distribution is sufficiently great that we prefer to use
this number to give an indication of the number of discrepant points
that it might be reasonable to consider eliminating in estimating the
scatter on the colors.

In some samples, prior to obtaining the fits, we have eliminated a very
small number (3-8\%) of galaxies with highly anomalous \VVII\ colors
but in all cases the number so deleted lies within the estimates of
field contamination. The bluest and faintest (\II $>22$) of these are
generally quite compact and may in fact be field HII galaxies, rather
than spheroidals. Obviously without spectroscopy of the \II $>22$
galaxies we cannot prove these anomalous objects are all field galaxies
and thus it is certainly conceivable that there exist {\it some} faint
spheroidal galaxies that are cluster members with anomalous colors.
Although retaining these would increase the scatter in our faintest
sample compared to the numbers we discuss below, their inclusion would
produce an extended tail to the color distribution around a narrow core
(as expected for field contaminants) and thus, as we seek a statistical
result, the uncertain identity of these objects does not seriously
affect our main conclusions.

\subsection{UV-Optical Color-Magnitude Relations} 

Figure~4 shows the color-magnitude relations for each of the three clusters
with different symbols for the important morphological types. The best
fit to the total spheroidal population is drawn as well as that derived
by BLE for their Coma E/S0 sample shifted to $z$=0.54 using the local
giant elliptical SED adopted earlier (see below for further discussion).

In fitting the color-magnitude relation we have, with the assistance of
Richard Bower, been able to use the same algorithm used by BLE in their
analysis of the local Coma+Virgo samples. The fit parameters
(A=intercept at \II=21, B=slope) and the quality in terms of the rms
residual for the various samples are given in Table~2.  The intercept
is evaluated at \II = 21 to simplify comparisons for samples with
different slopes.  Assuming a Gaussian distribution for the residuals
around the fit we expect fractional errors in the quoted rms of 15\%,
10\% and 7\% for sample sizes of 20, 50 and 100 galaxies respectively.  

Looking at the various fits, it can be seen that the {\it slope} of the
color-magnitude relation is relatively poorly constrained given the
narrow apparent magnitude range available. It is consequently difficult
to address the question of whether the three clusters are consistently
drawn from a population with a single slope. The different richnesses
of the three clusters is a further restriction in our analysis.
Cl0016+16 contains more spheroidals than the other two clusters
combined and thus dominates the statistics.

However, when we combine the samples for all three clusters, correcting
only for differential redshift and reddening effects discussed in \S3.1,
we find that the overall slope for the deepest sample of 141
morphologically-classed spheroidals (E, E/S0 and S0) to \II=23.0
is quite well-constrained ($-$0.070 $\pm$ 0.009). This value
is not significantly changed by restricting the fit to the  93 galaxies
to \II=22.0. Ignoring the very small mismatch between \VVII\ and rest-frame 
\UV\ so far as the {\it slope} of the relation is concerned, the BLE fit 
to the Coma data to the same luminosity limit yields a color-magnitude
slope of $-$0.082 $\pm$ 0.008, i.e.\ agreement to within $\sim 1 \sigma$. 
We thus see little evidence evolution in the slope of
the UV-optical color-magnitude relation for spheroidal galaxies 
out to $z=0.54$, in line with expectations if metallicity differences, 
rather than age, is the prime cause of color differences in these
distant cluster spheroids (Kodama \& Arimoto 1996).

If we adopt the slope observed for the combined sample and fit the
individual clusters to determine the variation in their intercepts we
find the values listed in Table~2.  These show a mean color of \VVII $=
2.31\pm 0.05$.  Correcting for the scatter between the clusters
expected from our calibration errors ($\pm 0.04$) this would indicate a
dispersion between the mean colors in the three clusters of $\ls 0.03$,
compared to the intrinsic scatter within each cluster, $\pm 0.07$ (see
below).  We conclude that the differences between the mean colors of
the spheroidal populations in these three clusters are certainly no
larger than the internal scatter within the clusters.  

The absolute comparison of the present-day color-magnitude relation in
the context of the HST data requires a careful treatment of the
mismatch between \VVII\ and rest-frame \UV\ and relies on our absolute
photometric calibration of WFPC-2 discussed above.  To make
predictions, we have used the differential color matching techniques
explained in detail by Arag\'on-Salamanca et al.\ (1993) and the
results are shown on Figures~4.  The predicted color-magnitude
relations indicate an \II = 21.0 galaxy in Cl0016+16 would have a color
of \VVII = $2.68^{+0.02}_{-0.03}$ (where the error includes a
contribution for the range in absolute luminosity in different
cosmologies).  The mean color observed for our clusters would thus
imply $\Delta\,\VVII = -0.37\pm 0.06$ from $z=0$--$0.54$.  While early
studies indicated little evolution in the colors of bright elliptical
galaxies out to $z\sim 0.2$--0.5 (e.g.\ Wilkinson \& Oke 1978;
Kristian, Sandage \& Westphal 1978), more
recent work has reversed this conclusion.  In particular our estimate
is in reasonable agreement with the precision measurement of
Arag\'on-Salamanca et al.\ (1993) for Cl0016+16 and Cl0054$-$27 (and
the previous results of Ellis et al.\ (1985) and Couch et al.\ (1985)
for these clusters), who find $\Delta\,\VI = -0.29\pm 0.06$.  The
bluing trend observed for our data out to $z=0.54$ is thus compatible
with these estimates which have been interpreted as supporting a high
redshift for the formation of the bulk of the stars in luminous cluster
ellipticals (Arag\'on-Salamanca et al.\ 1993).

Any detailed interpretation of the rest-frame UV--optical
color-magnitude relation at high redshift will unfortunately be
affected by the uncertain contribution of hot stars in later stage of
stellar evolution. Recent models (Yi 1996) suggest a
greater contribution to the $U$-band light from such phases than
assumed by BLE on the basis of Bruzual's (1983) models.  Although it
seems clear that post-main sequence produce the enhanced far-UV flux
seen in local ellipticals (Demarque \& Pinsonneault 1988; Dorman, Rood
\& O'Connell 1993), quantifying the effect depends on poorly-understood
parameters, including the composition and mass loss on the red giant
branch (Yi 1996). The important point to note here is that the observed
{\it scatter} around the C-M sequence, which is the principal
measurement of interest here, should be an upper limit to that arising
from main sequence stars.

\subsection{UV-optical Scatter and Morphological Variations}
 
We now turn to the prime motivation for the analysis, namely
constraining the scatter about the best-fit relationship both internally
within each cluster and for the combined sample after small corrections
for differential reddening and redshift effects.  As before, we have
chosen to follow BLE's prescription as precisely as possible. Table~2
summarises the various rms values as a function of cluster, limiting
magnitude and morphological class.

Only in Cl0016+16 is the sample large enough to attempt to measure the
scatter for the  Es and S0s classes separately. Regardless of sample
definition, the tightness of the C-M relation is remarkably small
($\leq 0.09$ mag). There is also no evidence that the scatter for S0s
is larger than that for those classed as Es.  The intrinsic
scatter in our sample can be obtained by subtracting the median
photometric errors assessed earlier as a function of apparent magnitude
in quadrature from the rms values measured to \II=21.0, 22.0, 23.0
(Table~2). This indicates the {\it intrinsic scatter} is
$\simeq$0.06$\pm$0.01 mag uniformly along the C-M sequence. Of the
other two clusters, the scatter in the E+S0 sample for Cl0054$-$27 is
comparable with that observed in Cl0016+16, while that in Cl0412$-$65
appears somewhat larger.  A bootstrap estimate of the error in the
Cl0412$-$65 measurement indicates an rms of $0.131\pm 0.027$ mag, so
the evidence for a larger scatter is marginal at best.

Having confirmed in the previous section that the scatter between the
clusters is smaller than their internal dispersion, we next combine the
data from the three clusters.  The enlarged sample enables us to
address possible morphological differences more carefully. As in
Cl0016+16, the S0 population presents a similar C-M relation and
scatter to that observed for the ellipticals.  In an attempt to
circumvent the problems with the misclassification of face-on S0's
discussed above we also compare the scatter for those E and S0's with
ellipticities above $\epsilon = 0.3$, where again we find that both
samples show comparable scatter, $\sim 0.07$ mag.  Although the
fraction of morphologically-classified 
S0s is small overall (a point we will return to later),
there is no convincing evidence for a broadening of their UV colours as
might be expected if a substantial fraction had been transformed from
spirals via environmental processes involving recent star formation.

In summary, the scatter of the rest-frame UV--optical C-M relation for
morphologically-confirmed spheroidal galaxies at $z$=0.54 is small. To
the same luminosity limit that BLE used in their analysis of the Virgo
and Coma clusters, the combined intrinsic scatter for E+S0s across all
three clusters is $<$0.07 mag, c.f.\ $<$0.035 locally (BLE), and the internal
scatter in each cluster is comparable. There is no evidence of a
systematic increase in the scatter when less luminous galaxies are
included (to $M_V=-17.8+ 5 \log h$) or between the E and S0
morphological samples, also in agreement with the analysis of local
clusters (Sandage \& Visvanathan 1978; BLE).

Finally, we have also looked for evolution of the colors {\it within}
the spheroidal galaxies in these clusters.  To do this we compare the
color within our 0.7 arcsec radius aperture with that measured in an
annulus between 0.7--1.5 arcsec radius ($\leq 11 h^{-1}$ kpc).  Taking
the 51 elliptical galaxies lying on the C-M sequences in the clusters
and for which we have good photometry (median \II = 21.1) we obtain a
median offset between the outer and inner regions of $\delta (V_{555} -
I_{814}) = -0.082\pm 0.037$, where the error is a bootstrap estimate of
the variance of the median.  For the average profile of our
ellipticals, these annuli correspond to luminosity-weighted radii of
0.18 arcsec and 0.96 arcsec respectively.  We therefore observe a color
gradient of $\delta (V_{555} - I_{814}) / \delta (\log r) = -0.11\pm
0.05$ per dex in these distant galaxies.  The color gradients observed
in local spheroidal galaxies are of the order of $-0.20 \pm 0.02$ in
$(U-R)$ and $-0.09\pm 0.02$ in $(B-R)$ (Peletier et al.\ 1990; Sandage
\& Visvanathan 1978).  Converting these to $(U-V)$ gives $\delta (U-V)
/ \delta (\log r) = -0.16\pm 0.03$, close to the value we obtain at
$z=0.54$.  This indicates that there also has been little relative color
evolution {\it within} these galaxies since $z\sim 0.5$, as expected
if these galaxies were all formed at high redshift.  

\section{Star Formation History of Cluster Spheroidals}

The tight scatter $\sigma{(V_{555}-I_{814})}$ we have found not only
internally in our three clusters but also across the entire sample
provides a significant new constraint on the history of star formation
in rich cluster galaxies. Returning to the analysis of BLE and
considering the close match between \VVII\ and \UV\ we can, to good
precision, apply their original formula based on rest-frame
$\delta$\UV:

$$\delta\UV = {d(U-V) \over dt} \, \beta (t_H - t_F) \leq \sigma$$

where ${d(U-V) / dt}$ is well-understood and governed solely by main
sequence lifetimes once the initial mass function is specified, $t_H$
and $t_F$ are respectively the cosmic age at the time of the
observations and the look-back time from then to the epoch at which
star formation in a single burst ended. $\beta$ is a factor of order
unity if the star formation history is uniformly distributed across the
interval between the big bang and $t_F$.  Values of $\beta$
significantly less than unity imply smaller ages for a given $\sigma$
but only at the expense of synchronising star formation across the
sample within the characteristic collapse timescale, $t_H - t_F$ at a
look-back time of $t_F$.

At low redshift, BLE grappled with separating $\beta$ and $t_F$ and
concluded by offering a joint constraint (Figure~5 of BLE). Assuming
$\sigma<$0.05 at $z\simeq$0 they placed the epoch of star formation as
long ago as 13 Gyr for $\beta$=1, but concluded lower ages of 6--10 Gyr
were allowed in the case where the spheroidals in Coma and Virgo had a
more synchronised star formation history corresponding to
$\beta$=0.1--0.3.

Whenever comparing ages  derived from main sequence lifetimes with
cosmological timescales the question of the value of the Hubble
Constant necessarily arises.  For the purposes of the discussion
presented below we will therefore adopt $h=0.5$ and note that using a
larger value for $h$ will tend to increase the redshift limits we
derive, although a high enough value (e.g.\ $h=1$) will lead to a
conflict between the age of the Universe and the stellar lifetimes
in standard cosmologies.

The most straightforward application of the BLE method is to assume the
scatter arises from the random formation of spheroids within a 1 Gyr
period which ended at $t_F$. The observed scatter of $<$0.07 mag gives
a minimum age of $t_F\simeq$5 Gyr {\it when viewed from a redshift of
0.54} irrespective of other assumptions.  A more elaborate treatment
invoking $\beta$=1 but based on a 1 Gyr burst in a cosmology with
$t_H$=15 Gyr as assumed by BLE ($h=0.5$), gives a similar minimum age of 6 Gyr
(Figure~6). As discussed in detail by BLE, as the rate of change of
\UV\ with time is governed primarily by the rate at which the
main-sequence turnoff evolves redward, such ages depend only weakly on
the slope of the IMF and upon metallicity (Iben \& Renzini 1984).

At high redshift, the ``$\beta$ problem'' is less important for two
reasons.  Firstly, we have witnessed that across {\it three} clusters
there is no convincing evidence that the cluster-cluster scatter is
significantly larger than that observed internally. Yet, it would
certainly appear that the three clusters are rather different in their
evolutionary histories, e.g.\ Cl0016+16 is much more massive, richer
and X-ray luminous than Cl0412$-$65.  It would be difficult to argue
that the rate of star formation in the respective cluster galaxies
could evolve at the same rate unless {\it either} spheroidal galaxies
evolve as closed boxes with no interference from their local
environment {\it or} the epoch of star formation was quite a long time
earlier.

Secondly, if the primary goal is to establish whether or not spheroidal
galaxies formed in a single burst at high redshift, the compression of
look-back time at high redshift means that $\beta$ uncertainties are
far less troublesome when interpreting the UV--optical scatter at $z$=0.54 as
compared to that viewed by BLE at $z$=0. For example, in a cosmology
with $h=0.7$ $\Omega_o=0.05$, an age of $>$6~Gyr at $z$=0.54
(corresponding to $\beta$=1) implies major star formation must have
occurred before $z_F\simeq 3$ when the age of the Universe was only
$\simeq$3 Gyr old. For $\Omega_o=1$, elliptical galaxies join the age
dilemma associated with globular clusters unless $H_o$ is
significantly lower.   The important point here is not the precise era of
formation but simply that {\it any} $\beta$ larger than 0.3 implies
cluster spheroidals formed most of their stars well before a redshift
of 3. Indeed, this era can only be brought {\it below} $z\simeq$3 if
the formation is unreasonably synchronised ($\beta\simeq$0.1) or by
invoking a world model with a dominant cosmological constant
($\Omega_{\Lambda}\simeq$0.8).

It is tempting to consider combining the BLE and present data into a
statement about the considerable age of {\it all} cluster spheroidals.
Certainly, BLE's estimates on the residual star formation allowed for
Virgo and Coma were {\it upper limits} entirely consistent with the new
constraints found here. However, as Kauffmann (1996) has argued, by
selecting the richest clusters at a given redshift for study, we are
unlikely to be observing the precursors of present-day clusters.
Likewise, Franx \& van Dokkum (1996) have warned of the selection
effects that might operate if ellipticals are identified in ways that
guarantee they are at least 2--3 Gyr old at any redshift.  It is
certainly possible that the luminous ellipticals in Coma and Virgo are
somewhat younger than those studied here and we cannot exclude the
possibility that more luminous ellipticals will form in, say, Cl0016+16
subsequent to the time of our observation. However, the latter does
seem unlikely given the elliptical fraction is already very much higher
than in many local clusters. The most robust statement we can make from
the present study is that the stars that form the dominant proportion
of the red light in three $z\sim 0.54$ clusters formed before
$z\simeq$3.

We have also shown that the differences in the mean colors of the
spheroidal galaxies in the three clusters are smaller than the
intrinsic dispersion within any one cluster ($\ls 0.03$ versus $\ls
0.07$).  The relatively wide range in cluster masses for our sample
would then indicate that spheroidal galaxies in rich clusters were {\it
all} formed within a relatively short period of time.  However, there
is expected to be some scatter between the collapse times of the
structures in which the ellipticals formed, at a fixed mass, depending
upon the final mass of the structure in which they are forming.  The
small cluster-to-cluster scatter would indicate that the environmental
variation in the mean collapse times is smaller than the intrinsic
dispersion in collapse times for halos, which will all form similar
luminosity galaxies. 

Two important questions emerge from the above discussion. Firstly, to
what extent can we generalise the conclusion concerning the bulk of the
starlight in these three clusters to the majority of rich clusters? It
certainly seems reasonable to consider that our conclusions might apply
to those galaxies that form the `red envelope' in the larger sample of 10
clusters tracked by Arag\'on-Salamanca et al.\ (1993) to $z\sim 0.9$.
Although those authors did not have access to morphological data, the
scatter in the $(V-K)$ color of the red envelope from cluster to cluster
at a given redshift is similarly small and the photometric color
evolution, when interpreted in the framework of Bruzual's (1983) models,
indicates a high redshift of star formation. To generalise our result
however, we would need to address the selection criteria for the
clusters themselves, as well as the volume density of red starlight
so located compared to the present-day value. As none of the clusters
in the present dataset, or in Arag\'on-Salamanca et al's study, were
found using well-defined search criteria, it is not possible to
make further progress.

The second question is to what extent we can quantify the proportion of
{\it present-day} spheroidals which have the properties we have assigned
to the sample seen at $z =0.54$? This is crucial to understand when one
considers recent attempts to track the fundamental plane to moderate
redshifts (van Dokkum \& Franx 1996; Barger et al.\ 1996b).  Although
related to the first question in the sense that any evolution of the
morphological content of clusters with redshift cannot be detached from
understanding how the clusters themselves were selected, we immediately
note from Table~2 that the E:S0 ratio in our high redshift clusters
is far higher than that observed in present-day clusters like Coma and
Virgo. Thus although we find no evidence that those few 
(predominantly luminous) S0s formed their
stars any later than the more dominant ellipticals, clearly the mixture
has evolved since $z=0.54$ in the sense that points to continued
production of S0s. This point is discussed more fully in the context of
the evolution of the morphology-density relation in the sample
discussed by Dressler et al.\ (1996).
 
If our results are typical of high redshift clusters, as seems
reasonable given earlier work, what can we expect to see during the
primeval star-forming phase? Traditionally, such systems were
thought to be spectacularly luminous placing them well within the
range of faint optical redshift surveys (Tinsley 1977). The absence of
a bimodal redshift distribution at faint limits (Lilly et al.\ 1995;
Cowie et al.\ 1996) makes this picture difficult to support unless the redshift 
of formation is very high. A more likely explanation is the hierarchical 
merging picture elucidated by Baugh et al.\ (1996) and Kauffmann (1996).
In this picture, a rapid era of star formation at $z>3$--4 involves
sub-units of lower intrinsic luminosity which subsequently merge and
assemble the galaxies seen at $z=0.54$. Kauffmann claims the merging 
history can be made consistent with our measured \VVII\ scatter if it 
occurs primarily before a redshift $z\simeq 2$ (c.f.\ her Figure~3),
at such an early epoch it would appear to be somewhat semantic whether
a halo collapse is described as a merger or not.

In recent months, the first glimpse of a possibly normal galaxy population at
redshifts $z\gs 3$ has emerged via Lyman limit selected samples in the
Hubble Deep Field (van den Bergh et al.\ 1996; Clements \& Couch 1996;
Giavalisco et al.\ 1996; Steidel et al.\ 1996b) and in other faint fields 
(Steidel et al.\ 1996a). The redshifts of many of these objects have now been
confirmed spectroscopically. Morphologically they are found to be small
objects; many are multiple and all are blue and undergoing moderate
star-formation.

It is tempting to interpret these high redshift systems as ancestors of
the spheroidal population. Crucial to this interpretation however, is an
estimate of the mass associated with the $z\simeq 3$ objects as well
as an estimate of their volume density in the context of the present-day
population. At this stage neither connection can be convincingly
established. However, we would note that the principle conclusion of
our study is that the main star formation era for cluster spheroidals
occurred before $z\simeq 3$, i.e.\ possibly in the regime of the
Lyman-limit samples and that these are most likely the oldest galaxies
of all. The connection with the Lyman limit samples is therefore certainly
suggestive when one considers the {\it rarity} of the population
observed beyond $z\simeq$3.

\section{Conclusions}
 
We have analysed the photometric properties of a large sample
of morphologically-selected spheroidal galaxies in three clusters of
mean redshift $z=0.54$ and shown how it is possible to derive 
constraints on the nature of the Universe beyond $z\simeq 3$ from 
precision studies at more modest redshifts. We can summarise our 
main conclusions as follows:

\begin{enumerate}

\item We demonstrate, through simulations and other tests that we can
easily differentiate between spheroidal galaxies and disk systems to at
least \II=22.0 in WFPC-2 images of a few orbits' duration. 

\item We find that the rest-frame UV-optical color-luminosity relation
has approximately the same slope at $z$=0.54 as it does locally with
only a modest bluing, $\Delta \simeq -0.3$ mag, consistent with
evolution inferred from earlier ground-based studies.  

\item The scatter in the rest-frame UV-optical color-luminosity relation 
at $z=0.54$ is found to be $\leq 0.1$ mag rms to \II=23.0 with no evidence 
of any luminosity-dependent scatter down to absolute magnitudes of 
$M_V=-17.8 + 5 \log h$. In the combined samples,
there is no evidence that the S0 population has a greater scatter
than the more numerous ellipticals, a similar result to that found
locally (Sandage \& Visvanathan 1978; BLE).

\item Significantly, the external scatter in the color-luminosity
relation between the three clusters is smaller than
the internal scatter within each cluster.   Given the relatively large
range in cluster properties within our sample, this would imply that
the formation epoch for the stellar populations in the cluster
spheroids is fairly insensitive to the cluster properties.
 
\item In the context of the earlier work of Bower, Lucey \& Ellis
(1992), we can understand the tight scatter in rest-frame UV-optical
colors only if the bulk of the star-formation in the dominant
spheroidal cluster galaxies occurred about 5--6 Gyr earlier than the
era at which they are observed. In conventional cosmologies without a
dominant $\Lambda$ term, this implies a redshift $z_F>3$ for $h=0.5$.
This high redshift is consistent with the modest bluing of their
stellar populations we observe between $z=0$ and $z=0.54$.

\item Although our result does not preclude the continued formation of
spheroidal galaxies in clusters or in the field at lower
redshift, within the context of hierarchical models, an important
implication of our result should be the detection of a population of
star-forming sub-units with $z>3$ as recently verified with Keck
spectroscopy.

\end{enumerate}

\section*{Acknowledgements}

We wish to thank Ray Lucas at STScI for his invaluable assistance which
enabled the efficient gathering of these observations.  We also thank
Barry Madore for generously providing the high quality ground-based
images of our clusters necessary for their calibration. We thank the
referee, Dr.\ Allan Sandage, for his clear and precise comments which
have helped clarify the structure of this paper.   We
acknowledge important contributions from Bob Abraham, Amy Barger,
Bianca Poggianti and, especially, Richard Bower whose methods we have
closely followed in this study. Finally we thank Alvio Renzini and
Michael Rich for their consistent encouragement.

%\vfil\eject

\smallskip

%\section*{Figure~Captions}
%\vfil\eject
%\centerline{\bf FIGURE CAPTIONS}

%\hbox{~}\vspace*{1truein}
\centerline{\hbox{
}}

\noindent{\bf Figure~1:} WFPC-2 images in the F814W band for three high
redshift clusters, (a) Cl0016+16 ($z=0.54$), (b) Cl0054$-$27 ($z=0.56$)
and (c) Cl0412$-$65 ($z=0.51$).  Objects marked represent
morphologically-selected early-type galaxies (E, E/S0 and S0) to \II
=23.0.
%\vfil\eject

%\hbox{~}\vspace*{1truein}
\centerline{\hbox{
}}
%\centerline{\bf Figure 1b}
%\vfil\eject

%\hbox{~}\vspace*{1truein}
\centerline{\hbox{
}}
%\centerline{\bf Figure 1c}
%\vfil\eject

\centerline{\hbox{
\psfig{figure=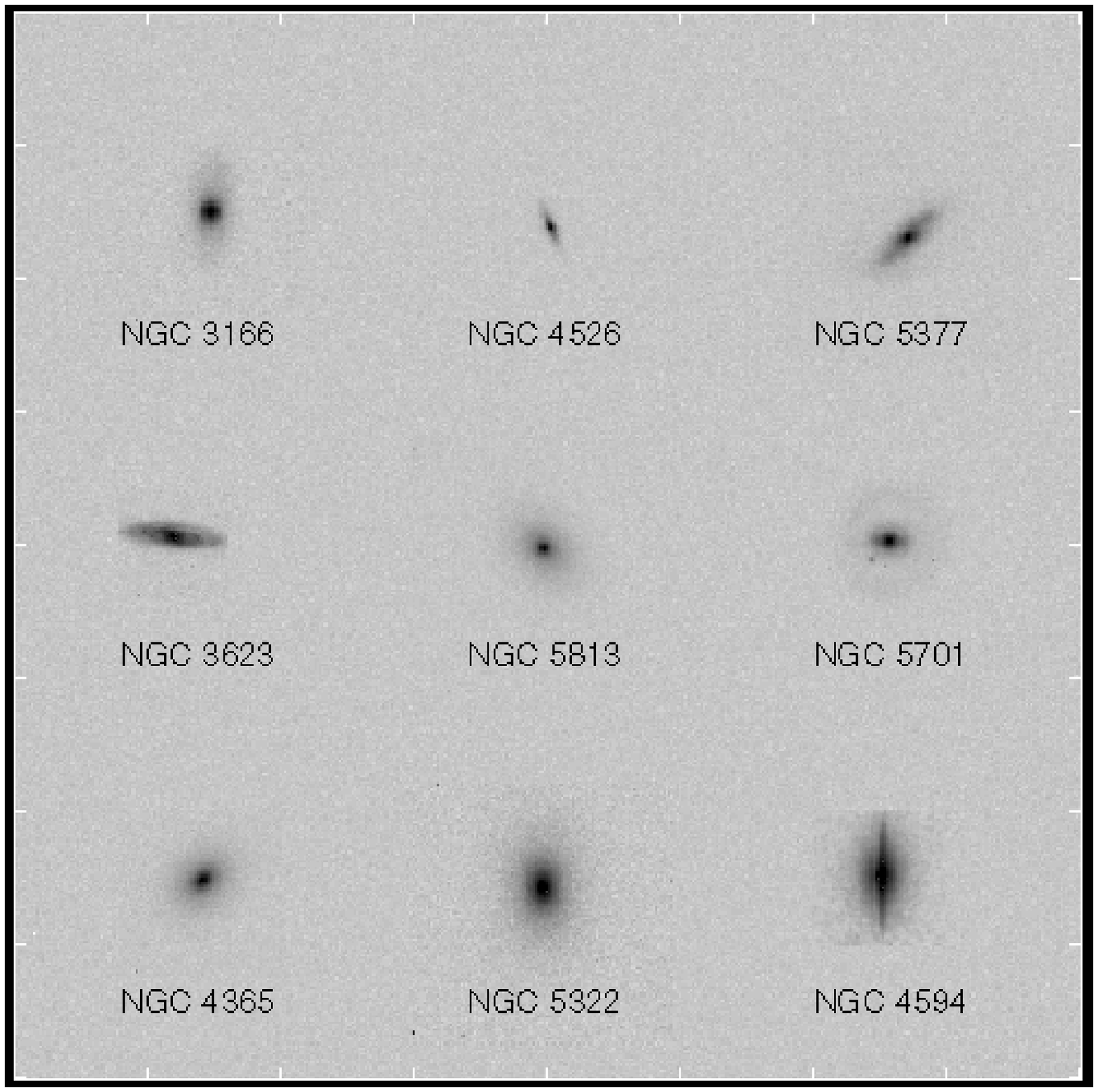}
}}

\noindent{\bf Figure~2:} The simulated appearance of representative
local E, S0 and early type spirals as they would appear when placed at
a redshift $z=0.5$ and observed with WFPC-2 using parameters equivalent
to the ``MORPHS'' survey. NGC 5813, 4365 and 5322 are Es, NGC 4526 is a
S0, NGC 5377 and 3623 are Sa's and NGC 4594, 5701 and 3166 are S0/a.
Classifications for these galaxies come from the RC3 (de Vaucouleurs et
al.\ 1991). The simulations were kindly performed by Roberto Abraham
and take into account pixel-by-pixel $k$-correction and surface
brightness effects with WFPC-2 specific parameters as discussed by
Abraham et al.\ (1996b).

\centerline{\hbox{
\psfig{figure=f3.ps,angle=270}
}}

\noindent{\bf Figure~3:}
A mosaic of F814W images and morphological classifications for
representative spheroidal galaxies in our three clusters. The top two
rows indicate galaxies in the bright sample \II\ $< 21.0$ with the
sub-class indicated according to a scheme E : E/S0 : S0 (see text).
The third row indicates objects classed as elliptical to the fainter
limit $21.0 < I_{814} <23.0$. The bottom row shows spheroidal galaxies for
which there is some disagreement amongst the classifiers (see text for
details). The individual images are $10 \times 10$ arcsec, roughly
equivalent to 40 h$^{-1} \times 40$ h$^{-1}$ kpc at the distance of our
clusters.

\centerline{\hbox{
\psfig{figure=f4a.ps}
}}
\noindent{\bf Figure~4}
The \VVII--\II\ color-magnitude diagrams for the three clusters after
transforming them to the Cl0016+16 observed frame, (a) Cl0016+16,
(b) Cl0054$-$27 and (c) Cl0412$-$65.
Es are indicated by filled circles, S0's by triangles and E/S0s by
squares. Those spheroidals and compact objects known to be field
galaxies or discounted from the analysis are indicated by open circles.
We overplot the best linear fits (Table~2) to the color-magnitude
relation of the spheroids (E, E/S0 and S0's) and the dashed line
represents the local BLE relationship as it would appear at $z$=0.54 in
the absence of any color evolution.
\vfil\eject

\centerline{\hbox{
\psfig{figure=f4b.ps}
}}
\centerline{\bf Figure 4b}
\vfil\eject

\centerline{\hbox{
\psfig{figure=f4c.ps}
}}
\centerline{\bf Figure 4c}
\vfil\eject

\centerline{\hbox{
\psfig{figure=f5a.ps}
}}
\noindent{\bf Figure~5:} 
(a) The distribution of \VVII\ color residuals for spheroids (E, E/S0
and S0's) in the combined cluster sample to \II=23.0 after removing the
mean color magnitude slope (see text for details). Light shaded objects
refer to those eliminated in the fits.  (b)  The similar distribution
for elliptical galaxies (E-only) from all three  clusters to a limit of
\II=22.0.  This has been corrected for the  mean color magnitude slope
(see text for details).  (c) The distribution of \VVII\ color residuals
for all the S0 galaxies brighter than \II=22.0 in the combined cluster
sample,  after correcting for the mean color magnitude slope. 

\vfil\eject
\centerline{\hbox{
\psfig{figure=f5b.ps}
}}
\centerline{\bf Figure 5b}
\vfil\eject

\centerline{\hbox{
\psfig{figure=f5c.ps}
}}
\centerline{\bf Figure 5c}
\vfil\eject

\centerline{\hbox{
\psfig{figure=f6.ps,height=5in,angle=270}
}}
\noindent{\bf Figure~6:}
The solid line shows the rate of change of galaxy color as a function
of time since the formation epoch $t_F$ after a single burst of star
formation of duration 1 Gyr. Assuming the observations refer to a
cosmic age $t_H$ associated with $z=0.54$, the observed scatter in
colors, $\sigma$, places an upper limit on the allowed rate of color
evolution according to the criterion: $\delta\UV /\delta t
\,<\,\sigma/\beta(t_H-t_F)$, where $\beta$ is a factor that allows any
synchronisation in the formation history (see text). The dotted lines
indicate minimum ages {\it at the time of observation} for various
values of $\beta$.

\vfil\eject

%\cleardoublepage
\singlespace
\begin{center}
\begin{tabular}{lcccccc}
\multispan7{\bf \hfil Table~1 \hfil }\\
\noalign{\medskip}
\multispan7{\bf \hfil Cluster Sample \hfil }\\
\noalign{\medskip}
\hline
\noalign{\smallskip}
{Cluster} & $\alpha$ & $\delta$ & $z$ & L$_X$ & \multispan2{T$_{\rm exp}$ (ks)} \\
\hfil & J2000 & J2000 & \hfil & 10$^{44}$ ergs sec$^{-1}$ &  F555W & F814W \\
\noalign{\smallskip}
\hline
\noalign{\smallskip}
Cl0016$+$16& $00^h 18^m 33\arcsper 64$ & $+16^\circ 25' 46\arcsper 1$ & 0.546 & 23.5 & 12.6 & 14.7 \\
Cl0054$-$27& $00^h 56^m 54\arcsper 59$ & $-27^\circ 40' 31\arcsper 3$ & 0.563 & 1.0 & 12.6 & 16.8 \\
Cl0412$-$65& $04^h 12^m 51\arcsper 65$ & $-65^\circ 50' 17\arcsper 5$ & 0.510 & 0.3 & 12.6 & 14.7 \\
\noalign{\smallskip}
\hline
\noalign{\bigskip}
\end{tabular}
\end{center}
\bigskip
\bigskip

\begin{center} 
\begin{tabular}{llcrllc}
\multispan7{\bf \hfil Table~2 \hfil }\\
\noalign{\medskip}
\multispan7{\bf \hfil Color-magnitude Results \hfil }\\
\noalign{\medskip}
\hline
% Use Clipped Least-Squares's fits
\noalign{\smallskip}
{\bf Cluster}&{\bf Sample}&{\bf $I_{814}^{lim}$}&{\bf N} & {\bf A($I_{814}=21$)} & {\bf \hfil B \hfil} & {\bf rms} \\
\noalign{\smallskip}
\hline
\noalign{\smallskip}
Cl0016$+$16 & E+S0 &$<$23 &91  &2.363(11)&$-$0.0673($\,$~99)&0.092 \\
\hfil     &        &$<$22 &57  &2.361(10)&$-$0.0677(144)&0.076 \\  
\hfil     &        &$<$21 &29  &2.339(18)&$-$0.0941(220)&0.063 \\ 
\hfil     &        &$<$23 &91  &2.364    &$-$0.0695     &0.092 \\ 
\noalign{\smallskip}
\hfil     & E      &$<$23 &48  &2.368(12)&$-$0.0605(126)&0.082 \\ 
\hfil     &        &$<$22 &36  &2.355(13)&$-$0.0857(167)&0.074 \\ 
\hfil     &        &$<$21 &19  &2.338(23)&$-$0.0981(248)&0.059 \\ 
\noalign{\smallskip}
\hfil     & S0     &$<$23 &15  &2.358(22)&$-$0.0549(263)&0.080 \\ 
\noalign{\smallskip}
Cl0054$-$27 & E+S0 &$<$23 &30  &2.270(18)&$-$0.0898(148)&0.091 \\
\hfil     &        &$<$22 &21  &2.252(21)&$-$0.1184(238)&0.081 \\ 
\hfil     &        &$<$23 &30  &2.265    &$-$0.0695     &0.096 \\ 
\noalign{\smallskip}
Cl0412$-$65 & E+S0 &$<$23 &20  &2.289(31)&$-$0.0906(287)&0.131 \\ 
\hfil     &        &$<$23 &20  &2.290    &$-$0.0695     &0.137 \\ 
\noalign{\smallskip}
All       & E+S0   &$<$23 &141 &2.333(10)&$-$0.0695($\,$~86)&0.108 \\
\hfil     &        &$<$22 &93  &2.337(10)&$-$0.0614(124)&0.091 \\
\hfil     &        &$<$21 &54  &2.342(18)&$-$0.0571(196)&0.081 \\
\noalign{\smallskip}
\hfil     & E      &$<$23 &65  &2.357(10)&$-$0.0618($\,$~99)&0.082 \\
\hfil     &        &$<$22 &48  &2.350(12)&$-$0.0708(137)&0.076 \\
\hfil     &        &$<$21 &27  &2.347(21)&$-$0.0748(197)&0.058 \\
\noalign{\smallskip}
\hfil     & S0     &$<$23 &22  &2.354(17)&$-$0.0558(214)&0.076 \\
\hfil     &        &$<$22 &19  &2.357(23)&$-$0.0468(366)&0.073 \\
\noalign{\smallskip}
\hline
\noalign{\bigskip}
\end{tabular}
\end{center}

\end{document}